\documentclass[fleqn,usenatbib]{mnras}
\usepackage[T1]{fontenc}
\usepackage{ae,aecompl}
\usepackage{graphicx}
\usepackage{amsmath}
\usepackage{amssymb}
\usepackage{txfonts}
\usepackage{cleveref}
\usepackage{bm}
\usepackage{hyperref}
\usepackage{tensor}
\usepackage{mathtools}

\usepackage{cleveref}
\usepackage{upgreek}
\usepackage{xcolor}

\graphicspath{./}

\newcommand{\ltsima}{$\; \buildrel < \over \sim \;$}
\newcommand{\lsim}{\lower.5ex\hbox{\ltsima}}
\newcommand{\gtsima}{$\; \buildrel > \over \sim \;$}
\newcommand{\gsim}{\lower.5ex\hbox{\gtsima}}

\title[Equivalence principle tests with FRBs]
{Consistent equivalence principle tests with fast radio bursts}
\author[Reischke, Hagstotz and Lilow]
{
Robert Reischke\thanks{E-mail:  \href{mailto:reischke@astro.ruhr-uni-bochum.de}{reischke@astro.ruhr-uni-bochum.de}}$^{1}$, Steffen Hagstotz\thanks{E-mail: \href{mailto:steffen.hagstotz@fysik.su.se}{steffen.hagstotz@fysik.su.se}}$^2$ and Robert Lilow\thanks{E-mail: \href{mailto:rlilow@campus.technion.ac.il}{rlilow@campus.technion.ac.il}}$^{3}$
\\
$^1$ Ruhr University Bochum, Faculty of Physics and Astronomy, Astronomical Institute (AIRUB),\\ \hspace{0.15cm} German Centre for Cosmological Lensing, 44780 Bochum, Germany
\\
$^2$ The Oskar Klein Centre for Cosmoparticle Physics,  Department of Physics, Stockholm University, Roslagstullsbacken 21A, SE-106 91 Stockholm, Sweden \\
$^3$ Department of Physics, Technion, Haifa 3200003, Israel
\\
}

\begin{document}

\pagerange{\pageref{firstpage}--\pageref{lastpage}}
\pubyear{2020}
\maketitle
\label{firstpage}

\begin{abstract}
Fast radio bursts (FRBs) are astrophysical transients of still debated origin. So far several hundred events have been detected, mostly at extragalactic distances, and this number is expected to grow significantly over the next years. The radio signals from the burst experience dispersion as they travel through the free electrons along the line-of-sight characterised by the dispersion measure (DM) of the radio pulse. In addition, each photon also experiences a gravitational Shapiro time delay while travelling through the potentials generated by the large-scale structure. If the equivalence principle (EP) holds, the Shapiro delay is the same for photons of all frequencies. In case the EP is broken, one would expect an additional dispersion to occur which could be either positive or negative for individual sources. Here we suggest to use angular statistics of the DM fluctuations to put constraints on the EP parametrized by the post-Newtonian parameter $\gamma$. Previous studies suffer from the problem that the gravitational potential responsible for the delay diverges in a cosmological setting, which our approach avoids. We carry out a forecast for a population of FRBs observable within the next years and show that any significant detection of the DM angular power spectrum will place constraints on the EP that are by a few orders of magnitude more stringent than current limits.
\end{abstract}


\section{Introduction}
\label{sec:intro}
Fast radio bursts (FRBs) are short transients lasting usually only a few milliseconds with frequencies ranging from $\sim 100$ MHz to several GHz. Due to the free electrons in the ionized intergalactic medium (IGM), the pulse experiences a dispersion $\Delta t \propto \nu^{-2}$, where the amplitude is called the dispersion measure (DM)  \citep[e.g.][]{thornton_population_2013, petroff_real-time_2015, connor_non-cosmological_2016, champion_five_2016,chatterjee_direct_2017} and is proportional to the integrated electron density along the line-of-sight. While the mechanism for the radio emission is still under debate,\footnote{For a compilation of currently proposed mechanisms for FRBs, see \href{https://frbtheorycat.org}{https://frbtheorycat.org} \citep{platts_living_2019}.} their isotropic occurrence and large observed DM suggest an extragalactic origin, so that the DM can be used to test the distribution of diffuse electrons in the large-scale structure (LSS). Several authors therefore proposed to use the DM inferred from FRBs as a cosmological probe, using either the averaged signal \citep{zhou_fast_2014,walters_future_2018} or the statistics of DM fluctuations \citep{masui_dispersion_2015,shirasaki_large-scale_2017,rafiei-ravandi_characterizing_2020,reischke_probing_2021,bhattacharya_fast_2020,takahashi_statistical_2021}.

Recently, FRBs have also been employed to test the equivalence principle (EP), which is one of the key axioms of General Relativity. If the EP holds, photons of different frequencies, as well as any other freely falling massless particles, should follow the same null-geodesic. Measuring any deviation from this would immediately indicate EP-breaking physics beyond the current standard model. There is a number of papers \citep[e.g.][]{krauss_test_1988,wei_testing_2015,wei_limits_2016,wei_tests_2016,wu_testing_2016,wu_new_2017,yang_testing_2017,yu_new_2018,xing_limits_2019,yao_testing_2020} dealing with this type of measurements, which all rely on measuring the Shapiro time delay \citep{shapiro_fourth_1964-1} for different messengers. The most recent ones focus on gravitational waves, gamma-ray bursts or FRBs. It is key to all Shapiro delay tests that the transient is short compared to the associated Shapiro delay, since this in principle determines the sensitivity of the measurement. All of the measurements listed above assume a metric with weak perturbations that vanish at infinity. In the parametrized post-Newtonian (PPN) formalism, this leads to the well known Shapiro delay equation:
\begin{equation}
\label{eq:shapiro_delay}
    t_\mathrm{grav} = -\frac{1+\gamma}{c^3} \int_{\boldsymbol{r}_\mathrm{e}}^{\boldsymbol{r}_\mathrm{o}} \mathrm{d}\lambda \;U\bigl(\boldsymbol{r}(\lambda)\bigr)\;,
\end{equation}
where the integral is performed from the source at $r_e$ to the observer at $r_o$. We introduced the PPN parameter $\gamma$ measuring deviations from the Newtonian time delay \citep{will_confrontation_2014}, $U$ denotes the Newtonian gravitational potential, and $c$ is the speed of light. For General Relativity one finds $\gamma = 1$, independent of frequency or particle type. Deviations from the EP can be parametrized in terms of the difference $\Delta \gamma$ between the $\gamma$ values of different messengers. \citet{wei_testing_2015}, for example, used FRBs to constrain $\Delta \gamma$ between photons of different frequencies.

\citet{minazzoli_shortcomings_2019} pointed out that the standard formulation is not well defined in a cosmological setting. Since the starting point is a perturbed Minkowski metric, \cref{eq:shapiro_delay} can only be applied to nearby sources. For corrections in a cosmological setting, also see \citet{nusser_testing_2016}. In addition, since the cosmic density field does not vanish at infinity, the resulting naive Shapiro delay in \cref{eq:shapiro_delay} diverges. It is usually argued that only taking into account contributions from some structures along the line-of-sight (and neglecting the remaining cosmic structure) leads to conservative limits, but \citet{minazzoli_shortcomings_2019} demonstrate that the neglected contributions can be positive or negative. Thus, any partial reconstruction of the density field is insufficient.

In this paper we demonstrate how FRBs can be used to set limits on EP breaking similar to the procedure suggested by \citet{nusser_testing_2016}, but by using statistical observables of the DM. This avoids dealing with unphysical quantities and the problematic boundary conditions for a cosmological setting. If the EP is violated, the observed DM consists of a term arising due to the dispersion induced by the free electrons in the LSS structure and and by a second term arising from the fact that the photons at two different frequencies experience a different Shapiro delay. Since FRBs are very short, any breaking of the EP would lead to a very strong deviation in the time delay due to the long paths involved in a cosmological setting. Therefore one can expect that any measurement of statistics of the DM at the expected level immediately rules out any substantial deviation from $\Delta \gamma = 0$.

The remainder of the paper is organized as follows: we review the current status of EP measurements using FRBs in \cref{sec:shapiro_Frbs} and highlight the associated problems. \Cref{sec:dispersion_measure_statistics} introduces the DM statistic and describes how a statistical analysis of FRBs removes the problem of the conventional analysis. We present the results in \cref{sec:results}. In \cref{sec:conclusions} we summarize our findings. Throughout the paper we will assume a $\Lambda$-cold-dark-matter ($\Lambda$CDM) scenario with fiducial parameters given by the best fit values of \citet{aghanim_planck_2020}.

\section{Shapiro delay tests of the equivalence principle}
\label{sec:shapiro_Frbs}
\subsection{Using FRBs to test the equivalence principle}
The observed time delay, $\Delta t_\mathrm{obs}$, between different frequency bands of an astrophysical transient can be split into several contributions:
\begin{equation}
    \Delta t_\mathrm{obs} = \Delta t_\mathrm{int} +  \Delta t_\mathrm{LIV} +  \Delta t_\mathrm{m} +  \Delta t_\mathrm{grav}\;.
\end{equation}
$ \Delta t_\mathrm{int}$ is the intrinsic time delay due to the source and the type of transient. In the case of FRBs this can be split into the DM contribution $ \Delta t_\mathrm{DM}$ and a potential source contribution $ \Delta t_\mathrm{s}$. The second term, $\Delta t_\mathrm{LIV}$, describes Lorentz invariance violation and is ignored here. The term $\Delta t_\mathrm{m}$ is a potential additional dispersion in case the photon is massive. From the relativistic dispersion relation
\begin{equation}
 E^2 = c^2 p^2 + m_\gamma^2 c^4
\end{equation}
one finds that massive photons of different frequencies propagate at different speeds. This also introduces a delay in the arrival time between different frequencies of the FRB signal:
\begin{equation}
 \Delta t_\mathrm{m}(z) = \left(\frac{m_\gamma c^2}{4 \pi \hbar} \right)^2 A(z) \left( \nu_1^{-2} - \nu_2^{-2} \right) \: ,
\end{equation}
with the redshift evolution given by
\begin{equation}
 A(z) = \int \mathrm d z' \frac{1}{(1+z')^2 H(z')}\;.
\end{equation}
Note that the frequency dependence has the same shape as the dispersion caused by the intergalactic plasma, so for any single source this would contribute to the total DM.

Both $\Delta t_\mathrm{LIV}$ and $\Delta t_\mathrm{m}$ produce non-zero effects on the cosmological background. As the current study will be only concerned with perturbations, these two contributions are ignored in what follows. With this we are left with
\begin{equation}
\label{eq:observed_time_delay}
 \Delta t_\mathrm{obs} = \Delta t_s + \Delta t_\mathrm{DM} +  \Delta t_\mathrm{grav}\;,    
\end{equation}
where the last term is the difference in the gravitational time delay, \cref{eq:shapiro_delay}, between photons of different frequency bands.

\subsection{Problems with cosmological tests}
The general problems with EP tests based on the Shapiro delay have already been pointed out by \citet{minazzoli_shortcomings_2019}, but we will briefly recap them here for the case of FRBs. So far, most of the cosmological tests consider a localised FRB (or any other object suitable for a differential Shapiro delay measurement as outlined in \cref{sec:intro}) and use the reconstructed potential, for example the one of the Milky Way, to calculate the predicted Shapiro delay. Since the Shapiro delay is usually much larger than the delay induced by the dispersion of the free electrons along the line-of-sight, any EP breaking ($\Delta\gamma \neq 0$) would completely dominate the time delay. It is then assumed that this gives a conservative bound on $\Delta\gamma$ since the Shapiro delay will just increase when adding more reconstructed data, i.e. another cluster along the line-of-sight.

The problem with this procedure is that the gravitational potential along the line-of-sight is influenced by all particles in the Universe. Thus, the Shapiro delay diverges if the density does not fall off. The reason for this problem is the choice of coordinates associated with \cref{eq:shapiro_delay}. At spatial infinity the potential and its derivative are expected to vanish. This boundary condition might be suitable for isolated objects, but it is not fulfilled in the cosmological setting with constant background density. In particular, \citet{minazzoli_shortcomings_2019} show that the cosmological part completely dominates the naive expectation of the Shapiro delay from any individual Keplerian potential, e.g. of the Milky Way.

Lastly, \citet{minazzoli_shortcomings_2019} show how to renormalise the divergence by choosing an appropriate time coordinate and derive an expression for the Shapiro delay which, however, is no longer monotonic with respect to the number of sources used to reconstruct the potential along the line-of-sight. Therefore one cannot place a conservative bound on $\Delta\gamma$, at least on cosmological scales, by just considering a subset of sources for which a reconstructed potential along the line-of-sight is readily available. 

\subsection{A way out}
In a cosmological setting, we can avoid the problem of a diverging time delay by considering a weakly perturbed Friedman-Robertson-Walker (FRW) line element in conformal Newtonian gauge within the PPN formulation:
\begin{equation}
    \label{eq:metric}
    \mathrm d s^2 = -\biggl(1+ \frac{2 \phi}{c^2} \biggr)  \,\mathrm c^2 d t^2 + a^2(t) \; \biggl(1 - \frac{2 \gamma \phi}{c^2} \biggr) \, \mathrm d \mathbf x^2
\end{equation}
with the gauge potential $\phi${, the scale factor $a$ and the comoving coordinates $\mathbf{x}$}. The corresponding time delay between photons of different frequencies can then be written as \citep{nusser_testing_2016}
\begin{equation} 
\label{eq:cosmo_shapiro}
  \Delta t_\mathrm{grav}(\hat{\boldsymbol{x}}) = \frac{\Delta\gamma}{c^3}\int^{\chi_\mathrm{s}}_{0} \mathrm{d}\chi\; a(\chi) \, \phi \big( \hat{\boldsymbol{x}}\chi,a(\chi) \big) \;,
\end{equation}
where $\chi$ is the comoving distance at the background level. This expression does not diverge in a cosmological setting since it respects the cosmological symmetry assumptions by construction. While this looks very similar to \cref{eq:shapiro_delay}, the perturbation $\phi$ is a Gaussian random field with zero mean, and therefore the time delay can acquire positive and negative contributions along the line-of-sight. 

Note that this expression is not suited for individual nearby lines-of-sight dominated by objects that cannot be described as small, linear perturbations to a FRW background. Furthermore, the absolute value of $\phi$ in any finite volume is not an observable quantity. However, these problems can be avoided by observing angular correlations of time delays. They are insensitive to individual nearby objects, and are independent of monopole contributions and thus the absolute value of $\phi$. The fact that local fluctuations of $\Delta t_\mathrm{grav}$ can be negative poses no problem since any EP breaking changes the expected correlations. Angular correlations can therefore be used to place an upper bound on the strength of EP violation.

\section{Dispersion measure and time delay statistics}
\label{sec:dispersion_measure_statistics}
The observed time delay in direction $\hat{\boldsymbol{x}}$ for a source at redshift $z$ is interpreted as a total DM via
\begin{equation}
\label{eq:delay_measure}
    \Delta t_\mathrm{obs}(\hat{\boldsymbol{x}},z) \propto \mathrm{DM}_\mathrm{tot}(\hat{\boldsymbol{x}}, z) \nu^{-2}\;.
\end{equation}
More explicitly, the time delay between two measured frequencies $\nu_{1,2}$ is
\begin{equation}
\begin{split}
    \Delta t_\mathrm{obs}(\hat{\boldsymbol{x}},z) = &\ t_{\nu_1,\mathrm{obs}}(\hat{\boldsymbol{x}},z) - t_{\nu_2,\mathrm{obs}}(\hat{\boldsymbol{x}},z) \\ = &\ \mathcal{K} \, \mathrm{DM}_\mathrm{tot}(\hat{\boldsymbol{x}},z) \, \left(\nu^{-2}_1-\nu^{-2}_2\right)\;,
\end{split}
    \end{equation}
where we absorb all the constants in $\mathcal{K} = e^2/(2\uppi m_e c)$. Here, $e$ and $m_e$ denote the charge and mass of an electron, respectively. The DM and time delays can therefore be used interchangeably. We will mostly work with the DM from now on since it is the quantity of interest for FRB correlation studies.

Any additional delay in the arrival time of different pulse frequencies due to EP breaking leads to a shift
\begin{equation}
\label{eq:dispersion_measure_observed}
    \mathrm{DM}_\mathrm{tot}(\hat{\boldsymbol{x}},z) \to \mathrm{DM}_\mathrm{tot}(\hat{\boldsymbol{x}},z) + \mathcal{D}_\mathrm{grav}(\hat{\boldsymbol{x}},z)\;,
\end{equation}
where $\mathcal{D}_\mathrm{grav}(\hat{\boldsymbol{x}},z)$ is the EP-breaking time delay from \cref{eq:cosmo_shapiro} interpreted as a DM in direction $\hat{\boldsymbol{x}}$ and up to redshift $z$:
 \begin{equation}
 \label{eq:dispersion_measure_grav}
    \mathcal{D}_\mathrm{grav}(\hat{\boldsymbol{x}},z) = \frac{\Delta\gamma}{\mathcal{K}c^3\left(\nu^{-2}_1-\nu^{-2}_2\right)}\int_0^{\chi(z)}\mathrm{d}\chi^\prime\; a(\chi^\prime) \, \phi\big(\hat{\boldsymbol{x}}\chi^\prime,z(\chi^\prime)\big)\;.
\end{equation}
As mentioned before, this contribution can be positive and negative, a problem which has not been addressed in previous studies. Note that this identification is subject to the $\nu^{-2}$ law, thus providing a preferred frequency shape of the EP-breaking term. However, the null hypothesis is $\Delta\gamma = 0$, as predicted by GR, and any additional contribution will immediately show up in the inferred DM budget. 

The non-gravitational contribution in \cref{eq:dispersion_measure_observed} is usually split into three parts:
\begin{equation}
\label{eq:dispersion_measure_contributions}
    \mathrm{DM}_\mathrm{tot}(\hat{\boldsymbol{x}},z) = \mathrm{DM}_\mathrm{LSS}(\hat{\boldsymbol{x}},z) + \mathrm{DM}_\mathrm{MW}(\hat{\boldsymbol{x}}) + \mathrm{DM}_\mathrm{host}(z)\;.
\end{equation}
$(i)$ For the contribution from the Milky Way, $\mathrm{DM}_\mathrm{MW}(\hat{\boldsymbol{x}})$, models of the galactic electron distribution predict $\mathrm{DM}_\mathrm{MW} \lsim 50 \; \mathrm{pc}\;\mathrm{cm}^{-3}$ \citep{yao_new_2017}. We will assume that this contribution can be modelled and subtracted from the signal. $(ii)$ For the host galaxy contribution, $\mathrm{DM}_\mathrm{host}(z)$, similar values are expected inducing a random scatter on the total DM. $(iii)$ The large-scale structure contribution can be written as
\begin{equation}
\label{eq:dispersion_measure_general}
    \mathrm{DM}_\mathrm{LSS}(\hat{\boldsymbol{x}},z) = \int_0^z n_\mathrm{e}({\boldsymbol{x}},z^\prime) \,     \frac{1+z^\prime}{H(z^\prime)}\mathrm{d}z^\prime\;.
\end{equation}
Here $H(z)$ is the Hubble expansion rate, and $n_e$ is the electron number density. The latter is a functional of the matter density contrast $\delta_\mathrm{m}(\boldsymbol{x},z)$:
\begin{equation}
\label{eq:electron_number_density}
    n_\mathrm{e}({\boldsymbol{x}},z) = \frac{\rho_\mathrm{b}({\boldsymbol{x}},z)}{m_\mathrm{p}} = \frac{\bar\rho_\mathrm{b}(z)}{m_\mathrm{p}}\left[1+b_\mathrm{e}({\boldsymbol{x}},z) \, \delta_\mathrm{m}({\boldsymbol{x}},z)\right] \; ,
\end{equation}
with the baryon mass density $\rho_\mathrm{b}({\boldsymbol{x}},z)$, its mean value $\bar\rho_\mathrm{b}(z)$, the proton mass $m_\mathrm{p}$ and the electron clustering bias $b_\mathrm{e}({\boldsymbol{x}},z)$. We assumed that the bias is linear, i.e. we consider only the lowest-order response of the electron density to the density of matter. This assumption should be valid on the scales we are considering. Rewriting \cref{eq:electron_number_density} yields:
\begin{equation}
\label{eq:dispersion_measure_specific}
    \mathrm{DM}_\mathrm{LSS}(\hat{\boldsymbol{x}},z) = \mathcal{A}\int_0^z \mathrm{d}z^\prime \frac{1+z^\prime}{E(z^\prime)}F(z^\prime)[1+b_\mathrm{e}({\boldsymbol{x}},z^\prime)\delta_\mathrm{m}({\boldsymbol{x}},z^\prime)]\;,
\end{equation}
with
\begin{equation}
    \mathcal{A} = \frac{3H_0^2\Omega_{\mathrm{b}0}\chi_H}{8\uppi G m_\mathrm{p}}\;,
\end{equation}
where $\chi_H$ is the Hubble radius today, $E(z) = H(z) / H_0$ the dimensionless expansion rate, and $F(z)$ the mass fraction of electrons in the IGM, which itself can be expressed as follows:
\begin{equation}
    \label{eq:ionization_fraction_of_the_igm}
    F(z) = f_\mathrm{IGM}(z)[Y_\mathrm{H}X_{\mathrm{e},\mathrm{H}}(z) + Y_\mathrm{He}X_{\mathrm{e},\mathrm{He}}(z)]\;.
\end{equation}
Here $Y_\mathrm{H} = 0.75$ and $Y_\mathrm{He} = 0.25$ are the mass fractions of hydrogen and helium, respectively, $X_{\mathrm{e},\mathrm{H}}(z)$ and $X_{\mathrm{e},\mathrm{He}}(z)$ are their ionization fractions, and $f_\mathrm{IGM}(z)$ is the fraction of electrons in the IGM. We assume $X_{\mathrm{e},\mathrm{H}} =X_{\mathrm{e},\mathrm{He}} = 1$ and $f_\mathrm{IGM}(z) = 90\%\; (80\%)$ at $z \gsim 1.5 \; (\lsim 0.4)$ \citep{meiksin_physics_2009,becker_detection_2011,shull_baryon_2012}.

To avoid dealing with the monopole, we study statistical properties of the DM. The angular power spectrum of the DM has been calculated in \citet{masui_dispersion_2015,rafiei-ravandi_characterizing_2020,reischke_probing_2021}. In the presence of a potentially EP-breaking time delay, the correlator of the observed DM fluctuations includes additional terms. We begin by splitting the LSS contribution to the DM into background and perturbation contributions:
\begin{equation}
\label{eq:fluctuation_definition}
     \mathrm{DM}_\mathrm{LSS}(\hat{\boldsymbol{x}},z) =  \overline{\mathrm{DM}}_\mathrm{LSS} (z) + \mathcal{D}_\mathrm{LSS}(\hat{\boldsymbol{x}},z)\,.
\end{equation}
$\mathcal{D}_\mathrm{LSS}(\hat{\boldsymbol{x}},z)$ is the effective DM induced by the fluctuations in the LSS and is given by the weighted line-of-sight integral over the electron density perturbation in the second term of \cref{eq:dispersion_measure_specific}. Note that there is no background contribution for the gravitational part, \cref{eq:dispersion_measure_grav}, which is exactly the problem in the ordinary measurements of EP breaking with FRBs. The perturbations to the DM from the electrons in the LSS and the Shapiro delay are therefore given by
\begin{equation}
\label{eq:curly_D_tot}
    \mathcal{D}(\hat{\boldsymbol{x}},z)  = \mathcal{D}_\mathrm{LSS}(\hat{\boldsymbol{x}},z)  + \mathcal{D}_\mathrm{grav}(\hat{\boldsymbol{x}},z) \;. 
\end{equation}

Given a normalised source redshift distribution $n(z)$, satisfying $\int\mathrm{d}z\; n(z) = 1$, and the associated distance distribution $n(\chi) = n(z) \, \mathrm{d}z / \mathrm{d}\chi$, \cref{eq:curly_D_tot} can be averaged over redshift:
\begin{equation}
\label{eq:dispersion_measure_averaged}
    \mathcal{D}(\hat{\boldsymbol{x}}) = \int_0^{\chi_H}\!\!\mathrm{d}\chi\; n(\chi) \, \mathcal{D}\big(\hat{\boldsymbol{x}},z(\chi)\big)\;. 
\end{equation}
By rearranging the integration limits we find
\begin{equation}
\label{eq:averaged_dispersion_measure_fluctuation}
    \mathcal{D}(\hat{\boldsymbol{x}}) = \int_0^{\chi_H}\!\!\mathrm{d}\chi\; W_\mathcal{D}(\hat{\boldsymbol{x}},\chi) \, \delta_\mathrm{m}
    \big(\hat{\boldsymbol{x}},z(\chi)\big) \;,
\end{equation}
\begin{figure}
    \centering
    \includegraphics[width=0.45\textwidth]{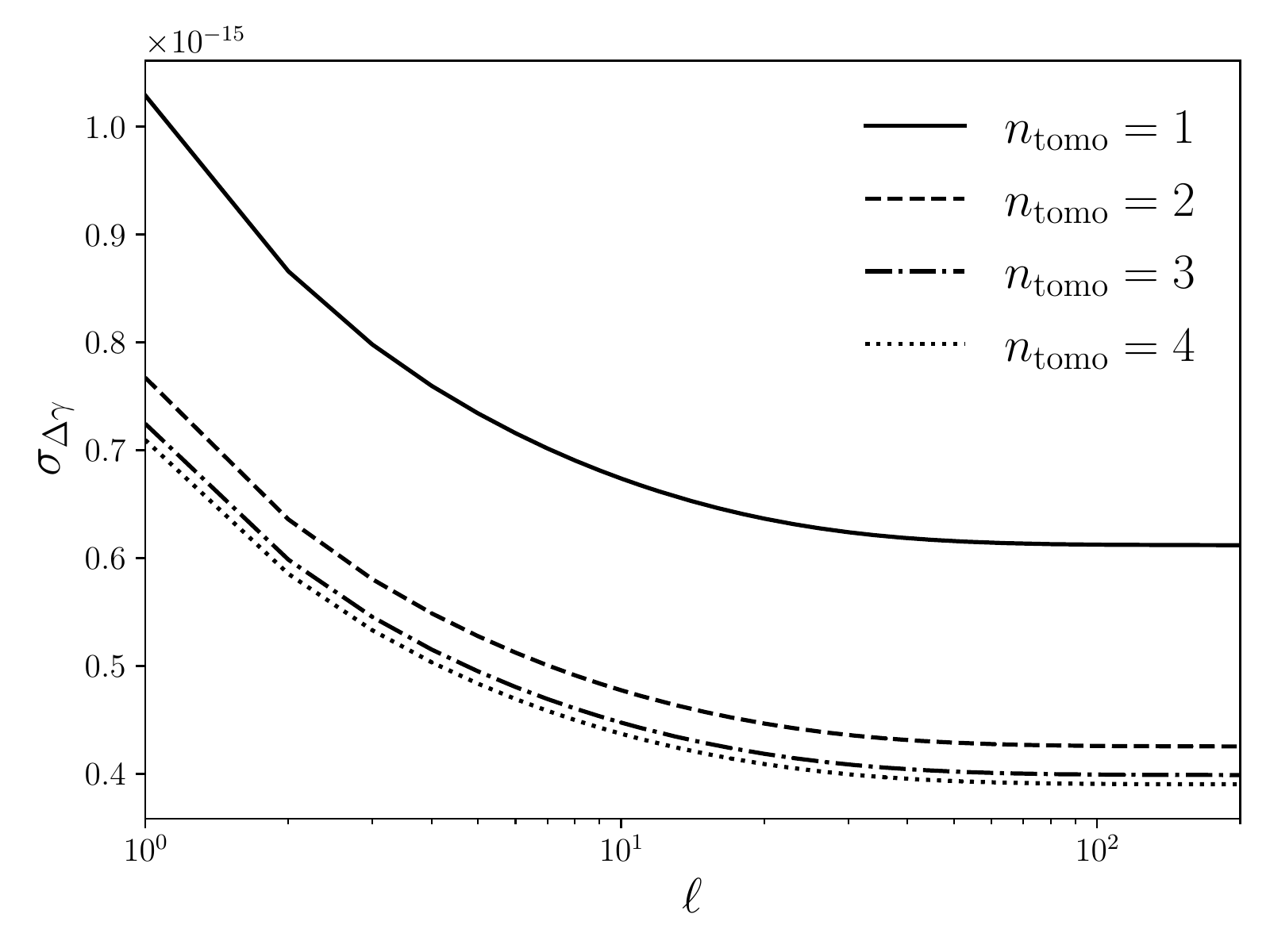}
    \caption{Possible constraints on the EP-breaking parameter $\Delta\gamma$ for a survey with $10^3$ FRB and $\alpha = 3.5$ and varying number of bins  $n_\mathrm{tomo}$ in the DM distribution.}
    \label{fig:constraints_delta_gamma}
\end{figure}
with the averaged weighting function
\begin{equation}
\label{eq:averaged_dispersion_measure_fluctuation_weighting_function}
    W_\mathcal{D}(\chi) = \bigg(W_\mathrm{LSS}(\hat{\boldsymbol{x}},\chi) + W_\mathrm{grav}(\hat{\boldsymbol{x}},\chi) \bigg) \int_\chi^{\chi_H} \mathrm{d}\chi^\prime n(\chi^\prime) \; .
\end{equation}
Here, $W_\mathrm{LSS}$ and $W_\mathrm{grav}$ are defined via \cref{eq:dispersion_measure_specific} and \cref{eq:dispersion_measure_grav}, respectively:
\begin{align}
\label{eq:averaged_dispersion_measure_fluctuation_weighting_function_LSS}
    W_\mathrm{LSS}(\hat{\boldsymbol{x}},\chi) &= \left|\frac{\mathrm{d}z}{\mathrm{d}\chi}\right|\mathcal{A} \, \frac{F \big(z(\chi) \big)  \, \big(1+z(\chi) \big)}{E\big(z(\chi)\big)} \, b_\mathrm{e}(\hat{\boldsymbol{x}},\chi) \,, \\
\label{eq:averaged_dispersion_measure_fluctuation_weighting_function_grav}
     W_\mathrm{grav}(\hat{\boldsymbol{x}},\chi) &=\frac{3\Omega_\mathrm{m0}\Delta\gamma\Delta^{-1}}{2\mathcal{K}c\chi_H^2\left(\nu^{-2}_1-\nu^{-2}_2\right)}\;,
\end{align}
where $\Delta^{-1}$ is the inverse Laplacian relating the potential fluctuations $\phi$ to the matter density contrast $\delta_\mathrm{m}$ via the Poisson equation. The angular power spectrum of DM correlations for the source distribution is then given by
\begin{equation}
\label{eq:dispersion_measure_angular_power_spectrum}
\begin{split}
    C^{\mathcal{D}\mathcal{D}}(\ell) &= \frac{2}{\uppi}\int\mathrm{d}\chi_1 \int\mathrm{d}\chi_2 
     \int k^2\mathrm{d}k\; W_\mathcal{D}(k,\chi_1) \, W_\mathcal{D}(k,\chi_2) \\  & \quad \times \sqrt{P_{\mathrm{mm}}(k,\chi_1) \, P_{\mathrm{mm}}(k,\chi_2)} \, j_{\ell}(k\chi_1) \, j_{\ell}(k\chi_2)\;,
    \end{split}
\end{equation}
where we express the weight function in Fourier space, so that the inverse Laplacian $\Delta^{-1} = k^{-2}$ for a comoving wavenumber $k$. The electron bias $b_\mathrm{e}$ is allowed to have a scale dependence.  Note that \cref{eq:dispersion_measure_angular_power_spectrum} depends on the two observed frequencies. Compared to the sensitivity to $\Delta\gamma$ this dependence is, however, very weak.

The host galaxy acts as a stochastic source with an intrinsic width of $\sigma^2_\mathrm{host}$. Due to the finite number of sources per solid angle, $\bar{n}$, this adds a white noise contribution to the observed spectrum:
\begin{equation}
\label{eq:dispersion_measure_angular_power_spectrum_observed}
    C^{\mathcal{D}\mathcal{D}}(\ell) \to C^{\mathcal{D}\mathcal{D}}(\ell)
    + \frac{\sigma^2_\mathrm{host}}{\bar{n}}\;.
\end{equation}
As shown in \citet{reischke_probing_2021}, the shot noise contribution is small compared to cosmic variance on large scales even if only a few thousand FRBs are available. It should also be noted that any redshift dependence of the host galaxy contribution to the DM has been averaged out, weighted by the source redshift distribution. We approximate this distribution by the following form:
\begin{equation}
\label{eq:source_redshift_distribution}
    n(z) \propto z^2\exp\left(-\alpha z\right)\;,
\end{equation}
where $\alpha$ determines the depth of the survey.

In \citet{reischke_probing_2021} the influence of the uncertainty on the DM in the absence of redshift information as well as its implication for the source redshift distribution and the corresponding angular power spectra was discussed. The DM is translated into a redshift assuming a fiducial cosmology without EP breaking. This can lead to dispersion space distortions \citep{masui_dispersion_2015} which we will, however, ignore here. However, the associated scatter of the redshift from the uncertainty of the DM is incorporated into the redshift distribution: \begin{equation}
\label{eq:observed_srd}
    n(z) = \int \mathrm d z_\mathrm{DM} \; n(z_\mathrm{DM}) p(z_\mathrm{DM} | z) \;,
\end{equation}
where $p(z_\mathrm{DM}|z)$ is well approximated by a Gaussian distribution, as shown in \citep{jaroszynski_fast_2019}:
\begin{equation}
\label{eq:photo_z_equi}
    p(z_\mathrm{DM}|z) \sim \mathcal{N}(\langle z_\mathrm{DM}\rangle (z),\sigma^2_{z_\mathrm{DM}(z)})
\end{equation}
with corresponding mean
\begin{equation}
\label{eq:mean_dispersion_measure}
    \langle z_\mathrm{DM}\rangle = \langle \mathrm{DM}_\mathrm{host} \rangle (z)+ \langle\mathrm{DM}_\mathrm{LSS}\rangle(z)
\end{equation}
and variance
\begin{equation}
\label{eq:sigma_dispersion_measure}
    \sigma^2_{z_\mathrm{DM}} = \sigma^2_\mathrm{host}(z) + \sigma^2_\mathrm{LSS}(z)\;.
\end{equation}
The latter consists of the host contribution $\sigma^2_\mathrm{host}$ and the cosmological contribution $\sigma^2_\mathrm{LSS}$.

We want to stress that any $\Delta\gamma \neq 0$ will introduce large fluctuations to the DM. Here we are analysing the data with respect to a $\Lambda$CDM scenario without EP breaking. In this sense we are assuming that we can convert a given DM into a corresponding redshift on the background level, while accounting for the scatter due to the perturbations in the electron density and the host contribution (see \citet{reischke_probing_2021} for more details).  \citet{rafiei-ravandi_characterizing_2020} discussed how cross-correlations could be used to further improve FRB distance measurements in the absence of redshift information.

Lastly it should be noted that everything could also be formulated in terms of the time delay directly, so that a time delay correlation function would be computed. This would be more closely related to the effects considered here. However, since FRB measurements are usually done in DM space, we stick to this notation here. 

\begin{figure}
    \centering
    \includegraphics[width=0.45\textwidth]{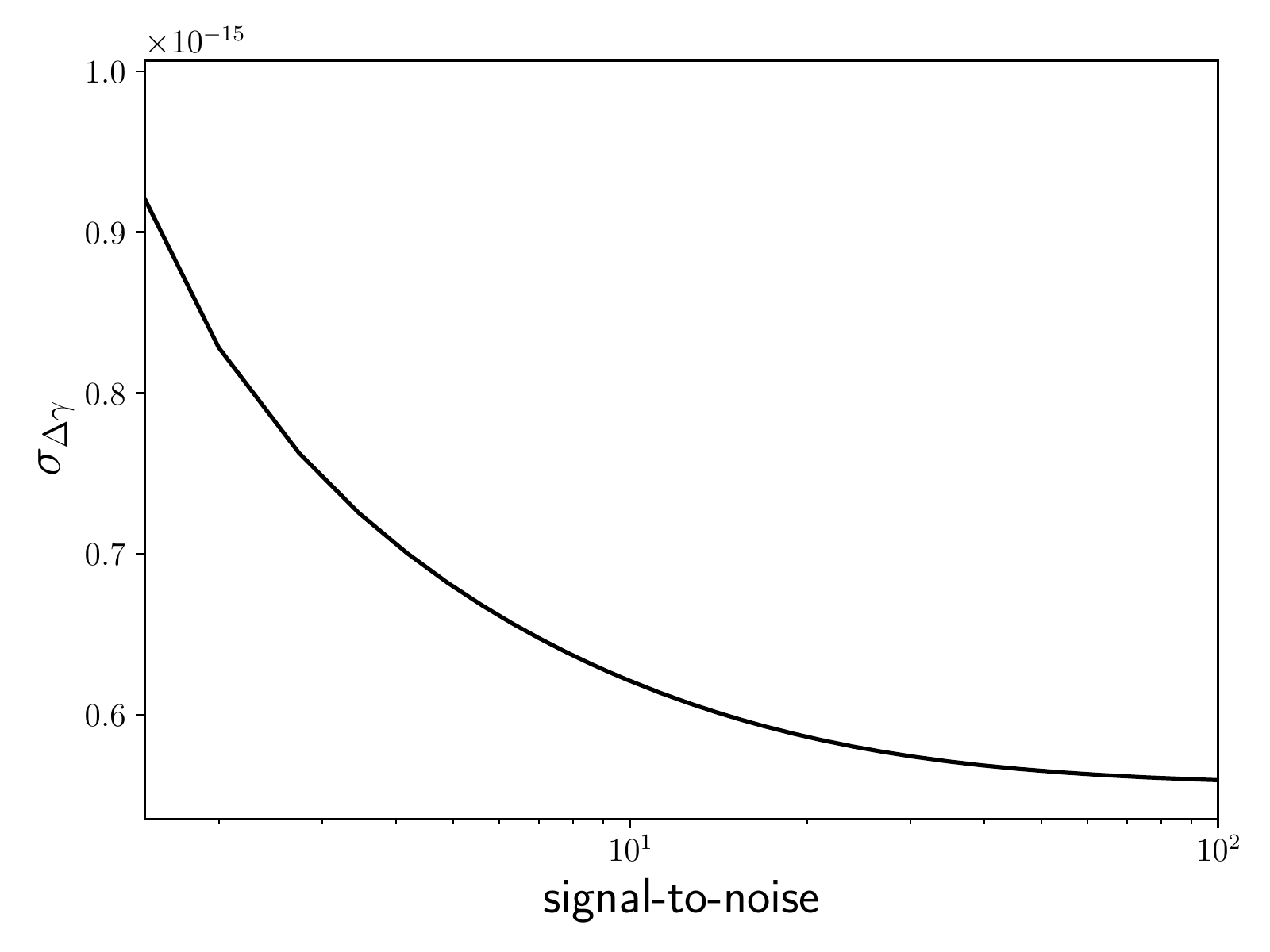}
    \caption{Constraints on the EP-breaking parameter $\Delta\gamma$ as a function of the signal-to-noise ratio of a measurement of the DM angular power spectrum.}
    \label{fig:constraints_delta_gamma_snr}
\end{figure}

\section{Results} 
\label{sec:results} 
In this section we will discuss our results for a forecast of an EP null test. For this we assume a survey following the source distribution \cref{eq:source_redshift_distribution} with $\alpha = 3.5$, which roughly corresponds to the distribution of currently observed FRBs \citep{petroff_frbcat_2016}. Note that we do not assume that the redshifts for the events are known, but only use the probabilistic conversion from DM to $z$ in \cref{eq:observed_srd}. Furthermore, we assume a total of $10^3$ observed FRBs. Lastly, the electron bias is assumed to be $b_\mathrm{e} = 0.8$ and it reaches unity at $z=3$, see \citep{reischke_probing_2021} for more details.

We consider an almost full sky survey, $f_\mathrm{sky}=0.8$, and decompose the DM map into spherical harmonics from which the angular power spectrum can be estimated. The sensitivity with respect to $\Delta\gamma$ is calculated using a Fisher forecast for a Gaussian likelihood of the spherical harmonic modes with covariance due to cosmic variance and Poisson noise. The latter describes the intrinsic DM scatter caused by the host and the finite number of FRB sources. Additionally, the DM maps can be separated into sub-samples by considering DM bins, as demonstrated in \citep{reischke_probing_2021}. This methods recovers some of the scale-dependent effects that are smoothed out by the line-of-sight projection. A pulse width of 0.3 GHz is assumed (corresponding to 1.5 and 1.2 Ghz), which enters in \cref{eq:averaged_dispersion_measure_fluctuation_weighting_function_grav}, and the host galaxy contribution is $\sigma_\mathrm{host}(z) = 50 \mathrm{pc} \, \mathrm{cm}^{-3} \, (1+z)^{-1}$.

In \cref{fig:constraints_delta_gamma} we show the resulting $1$-$\sigma$ limit on $\Delta\gamma$ for the described survey as a function of the maximum multipole considered and for different numbers of tomographic bins $n_\mathrm{tomo}$. The constraints presented by \citet{wei_testing_2015} are of the order of $10^{-10}$ while \citet{nusser_testing_2016} found $\sigma_{\Delta\gamma}\sim 10^{-13}$ for different FRBs. The constraints here are better by roughly two orders of magnitude. Since \citet{nusser_testing_2016} was interested in the rms-value and not the full correlation, the sensitivity to $\Delta\gamma$ is smaller than in the case studied here, as the angular power spectrum contains terms scaling quadratically with $\Delta\gamma$. The highest impact of the Shapiro delay measurements can be seen at low multipoles, while the constraining power settles very quickly at higher $\ell$. This happens for two reasons: $(i)$ the low multipoles are most affected due to the $k^{-2}$ factor in the Shapiro delay contribution coming from the Poisson equation. Therefore, the contribution of the gravitational time delay will dominate the DM most strongly on the largest scales. $(ii)$ Due to the small number of FRBs, the shot noise starts dominating over the signal at higher $\ell$. Increasing the number of tomographic bins increases the information content of the measurement to some degree, as it gets easier to pick up the scale-dependent imprint on the angular spectrum. However, the increased shot noise diminishes the gain when more than three bins are considered.

To present the constraints independent of the specific survey settings, we show in \cref{fig:constraints_delta_gamma_snr} again the $1$-$\sigma$ limit on $\Delta\gamma$, but now as a function of the signal-to-noise ratio of the measurement of the angular power spectrum. It is clear from the figure that any significant detection of the angular correlation in the DM of FRBs will immediately put very stringent constraints on deviations from the EP, $\Delta\gamma < 10^{-15}$, with increasing signal-to-noise only yielding mild improvements. The reason is that the integrated Shapiro delay over cosmological distances in combination with the short pulse durations yield $C^\mathcal{DD}_\mathrm{grav}$ contributions to the angular spectrum that are much larger than any dispersion caused by electrons.

For all forecasts made here we assumed all other parameters in the measurement to be fixed. Marginalizing over those, however, would not change the results dramatically since the effect of the Shapiro delay dominates the other contributions by many orders of magnitude if the EP is broken.
 
\section{Conclusions}
\label{sec:conclusions}
In this work we re-investigated constraints on the breaking of the equivalence principle (EP), characterized by the post-Newtonian parameter difference $\Delta \gamma$, with fast radio bursts by exploiting the Shapiro time delay. Previous studies had focused on individual FRBs with redshift information by accounting for the Keplerian potentials of known objects along the line-of-sight. This was shown to be inconsistent by \citep{minazzoli_shortcomings_2019}. Here we proposed instead to use statistical properties of the dispersion measure (DM) of FRBs to constrain deviations from the EP. This measurement does not suffer from divergences of the time delay and correctly accounts for the Shapiro delay induced by the LSS. In particular, we looked at the angular power spectrum of the DM of FRBs. While the Shapiro delay along any given line-of-sight can be positive or negative (causing problems with previous studies), in our approach the imprint on the angular DM correlations can always be detected as a change in the correlation structure. We performed a forecast for a null test of EP violation by considering a fiducial $\Lambda$CDM cosmology and placing upper limits on $\Delta\gamma$ between observed frequencies. We would like to stress that the present parametrization is just a phenomenological one without a direct link to any specific EP-breaking theory. It is in this sense just an agnostic test of the degree up to which the EP holds between two frequencies. We summarize our principal results as follows:

    $(i)$ We calculated the contribution to the DM angular correlation function from the free electron distribution in the LSS and the induced Shapiro delay by the LSS.
    
    $(ii)$ Using correlations of the DM measured from an FRB population avoids the usual problems with Shapiro delay measurements by satisfying appropriate boundary conditions in a cosmological setting. 
    
    $(iii)$ Detecting the angular power spectrum of the FRB DM with any significance at the predicted level will immediately place tight constraints on $\Delta\gamma$, which are a few orders of magnitudes better than the current limits.  

While we presented the theoretical predictions in DM space, one could alternatively formulate everything in time delay space, providing a closer connection to the Shapiro delay. Since the measurements of the angular correlation function will happen on very larges scales, due to the relatively high shot noise contribution from the limited number of expected FRB detections, it will be necessary to calculate general relativistic projection effects to the angular power spectrum to make accurate theoretical predictions. We intend to do this in a future paper. Due to the high sensitivity on $\Delta\gamma$ this will, however, not change our main conclusion that EP can be tested to unprecedented precision once the angular correlation of the FRB's DM has been detected.

\section*{Acknowledgments}
RR is supported by the European Research Council (Grant No. 770935). SH acknowledges support from the Vetenskapsr\r{a}det (Swedish Research Council) through contract No.\ 638-2013-8993 and the Oskar Klein Centre for Cosmoparticle Physics. RL acknowledges support by a Technion fellowship.

\label{lastpage}
\bibliographystyle{mnras}
\bibliography{MyLibrary}
\end{document}